\documentclass[pra,twocolumn,showpacs,preprintnumbers,amsmath,amssymb,superscriptaddress]{revtex4}

\usepackage{graphicx}% Include figure files
\usepackage{dcolumn}% Align table columns on decimal point
\usepackage{bm}% bold math%\documentclass[aps,prl,twocolumn,groupedaddress]{revtex4}
\usepackage[dvips]{epsfig}

\begin{document}

%\preprint{}


\title{Electric Field effects on quantum correlations in semiconductor quantum dots}
\author{S.~Shojaei}
\altaffiliation{%
Author to whom correspondence should be addressed; electronic
mail: shojaei.sh@gmail.com, s-shojaei@tabrizu.ac.ir}

\affiliation{%
Photonics Group, Research Institute for Applied Physics and Astronomy (RIAPA), University of Tabriz, 51665-163 Tabriz, Iran}

\author{M.~Mahdian}

\affiliation{%
Faculty of Physics, Theoretical and astrophysics department , University of Tabriz, 51665-163 Tabriz, Iran}

\author{R.~Yousefjani}

\affiliation{%
Faculty of Physics, Theoretical and astrophysics department , University of Tabriz, 51665-163 Tabriz, Iran}
%


%\date{\today}

\begin{abstract}

 We study the effect of external electric bias on the quantum correlations in the array of optically excited coupled semiconductor quantum dots. The correlations are characterized by the quantum discord and concurrence and are observed using excitonic qubits. We employ the lower bound of concurrence for thermal density matrix at different temperatures. The effect of the F\"{o}rster interaction on correlations will be studied. Our theoretical model detects nonvanishing quantum discord when the electric field is on while concurrence dies ,ensuring the existence of nonclassical correlations as measured by the quantum discord.

\end{abstract}
%


%\keywords{Biexciton, Binding energy, Built-in electric field, Two photon absorption}


\maketitle

\section*{1 Introduction}

In the recent years, the meaning of  quantum discords has been attracted a lot of interest because it indicates that entangled states are not the only kind of quantum states exhibiting nonclassical features\cite{zurek, Zurek2, Henderson, Luo, Vedral, Datta}. Quantum discord is a measure of the discrepancy between two natural yet different quantum analogs of the classical mutual information . Quantum discord captures the fundamental feature of the quantumness of nonclassical correlations, which is much like quantum entanglement, but it is beyond quantum entanglement because quantum discord is even present in separable quantum states. Quantum discords have strong potentials in studying dynamical processes\cite{zurek3, oppenheim, Horodecki, mahdian}, some quantum information processes such as the broadcasting of quantum states\cite{piani1, piani} quantum state merging \cite{cavalcanti, madhok}, quantum entanglement distillation\cite{cornelio, piani3} ,
entanglement of formation \cite{fanchini}.\\
Besides, there has been considerable interest in the quantum information properties of solid state environment particularly, semiconductor quantum dots (QDs) due to their well defined controllable atom-like and molecule-like properties\cite{filippo, filippo2,hawrylak}. To study the electro-optical properties of  QDs (specially coupled ones), one of the most interesting parameter is the exciton-exciton interaction. This kind of interaction in first neighbors dots will allow to implement a scheme for quantum information processing on QDs \cite{chen}. It was shown that the biexcitonic shift due to the dipolar interaction allows for subpicosecond quantum gate operations\cite{rinaldis02}.
It was proven that the optically driven QDs, that are carrying excitons are good candidates for  implementation of quantum gates and quantum computation\cite{xiaoqin,Herschbach,Friedrich}.

In order to study the amount of concurrence and discord in the array of excitons inside QDs, we employ the lower bound of concurrence for thermal density matrix of identical and equidistant coupled QDs at different bath temperatures. Furthermore, By the means of electric field, the manipulation of entanglement and discord will be discussed. We show that, quantum correlations may be enhanced upon increasing the temperature and decreased with incresing the electric field across the system. Our theoretical model detects nonvanishing quantum discord, ensuring the existence of nonclassical correlations as measured by the quantum discord

\section{2 Theory: quantum dot model and Hamiltonian}

The model sample to study the quantum correlation properties of  optically driven QDs is a series of InAs coupled QDs with small equal spacing between them along the axis. This model can be realized experimentally(see for example reference\cite{Nishibayashi}). In this model, F\"{o}rster mechanism\cite{Forster} is a valid model to explain the energy transfer between QDs through dipolar interaction between the excitons. Here, the qubits are the excitonic electric dipole moments located in each QD which can only orient along ($|0\rangle$) or against ($|1\rangle$) the external electric fields. For such a system, tuning and controlling the quantum correlation between the dipoles is of great importance. The governing Hamiltonian of dipoles in the presence of external electric field simply reads:

\begin{eqnarray}
H = \hbar\sum _{i=1}^{n}\omega_i[S_z^i+\frac{1}{2}]+\nonumber\\
+\hbar\sum_{i=1}^{n}\Omega_i\hat{S_i^z}+\hbar\sum_{i=1}^{n}J_z[S_+^i][S_-^j]+\nonumber\\
\frac{1}{2}\sum_{i,j=1}^{n}\lambda[S_+^iS_-^j+S_-^jS_+^i],
\end{eqnarray}

Where $S_+^i$=$(|0\rangle\langle1\rangle)$,$S_-^i$=$(|1\rangle\langle0\rangle)$, and $S_z^i$=$\frac{1}{2}(|0\rangle\langle0\rangle-|1\rangle\langle1\rangle)$. $\omega_i$ presents the frequency of the excitons in QDs, $\Omega_i$ is the frequency related to dipole moment(exciton) that is a function of dipole moment and the external electric field (E) at \textit{i}th QD:

\begin{equation}
\hbar\Omega_i=|\vec{d}.\vec{E}|,
\end{equation}

Where $\vec{d}$ is the electric dipole moment carried by the exciton that is assumed to be same for each QD. $\lambda$ presents the F\"{o}rster interaction which transfers an exciton from one QD to other ones. $J_z$ presents the exciton-exciton dipolar interaction energy, reads:

\begin{eqnarray}
\hbar J_z=\frac{\vec{d}^2(1-3\cos^2\theta)}{\vec{r}_{ij}^3},
\end{eqnarray}

where $r_{ij}$ is the distance between dipoles \textit{i} and \textit{j} that is assumed along the $z$ axis.

For a qualitative discussion on quantum correlations we assume that the dipole carried by exciton is the one order of magnitude in debyes, a typical
experimental electric field $10^6$ $V/m$, so the dipolar interaction parameter($\hbar J_z$) will be of the order of $meV$. F\"{o}rster interaction energy($\hbar \lambda$) and $\hbar \Omega$ are assumed to be in the order of $meV$. This values are consistent with experimental observations and calculations\cite{zhu}.

\subsection{2.1 Lower bound concurrence}

Since entanglement is conceived as a resource to perform various tasks of quantum information processing \cite{23,24,25,26}, knowledge about the amount of entanglement in a quantum state is so important. Indeed, awareness from the value of entanglement, means knowing how well a certain task can be accomplished. The quantification problem of entanglement only for bipartite systems in pure states \cite{27} and two-qubit system in mixed state \cite{28} is essentially solved. In multi-partite systems, even the pure state case, this problem is not exactly solved and just lower bounds for the entanglement have been proposed \cite{29,30,31,32}. Here, in order to determination the exact minimum of entanglement between three dipoles(excitons), we use the lower bound of concurrence for three-qubit state which is recently suggested by Li et al. \cite{33}
\begin{eqnarray}
\tau_{3}(\rho)=\frac{1}{\sqrt{3}}(\sum_{j=1}^{6}(C_{j}^{12|3})^{2}+(C_{j}^{13|2})^{2}+(C_{j}^{23|1})^{2})^{\frac{1}{2}},
\end{eqnarray}
where $C_{j}^{12|3}$ is terms of the bipartite concurrences for qubits $12$ and $3$ which is given by
\begin{eqnarray}
C_{j}^{12|3}=\max \{0,\lambda_{j}^{12|3}(1)-\lambda_{j}^{12|3}(2)-\lambda_{j}^{12|3}(3)-\lambda_{j}^{12|3}\}.
\end{eqnarray}
In this notation, $\lambda_{j}^{12|3}(\kappa)$, $(\kappa=1..4)$, are the square nonzero roots, in decreasing order, of the non-Hermitian matrix $\rho \tilde{\rho}_{j}^{12|3}$. The matrix $\tilde{\rho}_{j}^{12|3}$ are obtained from rotated the complex conjugate of density operator, $\rho^{*}$, by the operator $S_{j}^{12|3}$ as $\tilde{\rho}_{j}^{12|3}=S_{j}^{12|3}\, \rho^{*} \, S_{j}^{12|3} $. The rotation operators $S_{j}^{12|3}$ are given by tensor product of the six generators of the group SO(4), ($L_{j}^{12}$), and the single generator of the group SO(2), ($L_{0}^{3}$) that is $S_{j}^{12|3}=L_{j}^{12} \otimes L_{0}^{3}$. Since the matrix $S_{j}^{12|3}$ has four rows and columns which are identically zero, so the rank of non-Hermitian matrix $\rho \tilde{\rho}_{j}^{12|3}$ can not be larger than 4, i.e., $\lambda_{j}^{12|3}(\kappa)=0$ for $\kappa\geq 5$. The bipartite concurrences $C^{13|2}$ and $C^{23|1}$ are defined in a similar way to $C^{12|3}$.

In order to calculate thermal entanglement, we need the temperature dependent density matrix and the density matrix
for a system in equilibrium at a temperature T reads: $\rho=\exp(-\beta\hat{H}/Z)$ with $\beta=1/KT$ and $Z$ is partition function, $Z=Tr(\exp(-\beta\hat{H}))$. In this case, the partition function is

\begin{eqnarray}
Z(T)=\sum_{i=1}g_ie^{-\beta\lambda_i},
\end{eqnarray}

where $\lambda_i$ is the $i$th eigenvalue and $g_i$ is the degeneracy. and the corresponding density matrix can be written

\begin{eqnarray}
\rho(T)=\frac{1}{Z}\sum_i^N e^{-\beta\lambda_i}|\Phi_i\rangle\langle\Phi_i|,
\end{eqnarray}

here $|\Phi_i\rangle$ is the $i$th eigenfunction. The density matrix for our considered system has the form as:

\begin{equation}
 \rho(T) = \left( \begin{array}{cccccccc}
\rho_{11}&0&0&0&0&0&0&0 \\  0&\rho_{22}&\rho_{23}&0&\rho_{23}&0&0&0 \\ 0&\rho_{23}&\rho_{22}&0&\rho_{23}&0&0&0 \\ 0&0&0&\rho_{44}&0&\rho_{46}&\rho_{46}&0 \\ 0&\rho_{23}&\rho_{23}&0&\rho_{22}&0&0&0 \\ 0&0&0&\rho_{46}&0&\rho_{44}&\rho_{46}&0 \\ 0&0&0&\rho_{46}&0&\rho_{46}&\rho_{44}&0 \\ 0&0&0&0&0&0&0&\rho_{88}\end{array} \right),
\end{equation}

with
\begin{eqnarray}
\rho_{11}&=&\frac{e^{\beta\lambda}}{1+e^{\beta(2d+Fz+w)}}
\cr\cr&\times&[e^{\beta(2d+2Fz+w)}+e^{\beta\lambda}-e^{\beta(2d+Fz+w+\lambda)}\cr&&
+e^{\beta(4d+2Fz+2w+\lambda)}+2e^{\beta(2d+2Fz+w+\frac{3}{2}\lambda)}]^{-1},
\cr\cr\cr
\rho_{22}&=&\frac{1}{3}(1+2e^{\frac{3}{2}\beta\lambda})\cr&\times&
[(1+e^{\beta(2d+Fz+w)})(1+2e^{\frac{3}{2}\beta\lambda})\cr&&
+ e^{\beta(\lambda-2d-2Fz-w)}+e^{\beta(4d+Fz+2w+\lambda)}]^{-1},
\cr\cr\cr
\rho_{23}&=&\frac{e^{\beta(2d+2Fz+w)}(1-e^{\frac{3}{2}\beta\lambda})}{3(1+e^{\beta(2d+Fz+w)})}
\cr\cr&\times&
[e^{\beta(2d+2Fz+w)}+e^{\beta\lambda}-e^{\beta(2d+Fz+w+\lambda)}\cr&&
+e^{\beta(4d+2Fz+2w+\lambda)}+2e^{\beta(2d+2Fz+w+\frac{3}{2}\lambda)}]^{-1},
\cr\cr\cr
\rho_{44}&=&\frac{1}{3}(1+2e^{\frac{3}{2}\beta\lambda})\cr&\times&
[(1+e^{-\beta(2d+Fz+w)})(1+2e^{\frac{3}{2}\beta\lambda})\cr&&
+ e^{\beta(\lambda-4d-3Fz-2w)}+e^{\beta(2d+w+\lambda)}]^{-1},
\cr\cr\cr
\rho_{46}&=&\frac{e^{\beta(4d+3Fz+2w)}(1-e^{\frac{3}{2}\beta\lambda})}{3(1+e^{\beta(2d+Fz+w)})}
\cr\cr&\times&
[e^{\beta(2d+2Fz+w)}+e^{\beta\lambda}-e^{\beta(2d+Fz+w+\lambda)}\cr&&
+e^{\beta(4d+2Fz+2w+\lambda)}+2e^{\beta(2d+2Fz+w+\frac{3}{2}\lambda)}]^{-1},
\cr\cr\cr
\rho_{88}&=&e^{3\beta d}
\cr\cr&\times&[e^{3\beta d}+e^{-3\beta(d+Fz+w)}+2e^{\beta(d-w+\frac{1}{2}\lambda)}\cr&&
+e^{\beta(d-w-\lambda)}+e^{-\beta(d+Fz+2w+\lambda)}\cr
&&+2e^{-\beta(d+Fz+2w+\frac{1}{2}\lambda)}]^{-1}.
\end{eqnarray}

Since the explicit expressions of solutions of Eq. (7) are very complicated, here we skip the details and give our results in terms of figures. The discussion of results will be postponed to the next section.

\subsection{2.2 Quantum discord}

For a bipartite system AB quantum discord is defined by the discrepancy between quantum versions of two classically
equivalent expressions for mutual informationis \cite{20}
\begin{eqnarray} \label{2-2}
\textit{D}(\rho^{AB})=\textit{I}(\rho^{AB})- \textit{C}(\rho^{AB}),
\end{eqnarray}

where $\textit{I}(\rho^{AB})=S(\rho^{A})+S(\rho^{B})-S(\rho^{AB})$ and the classical correlation is the maximum information about one subsystem $\rho^{A}$ or $\rho^{B}$ which depends on the type of measurement performed on the other subsystem such as $\textit{C}(\rho^{AB})=max[S(\rho^{A})-S(\rho^{AB}|\{\Pi_{k}\})]$ with $S(\rho)=-Tr[\rho \log_{2}\rho]$ as the von-Neumann entropy. Notice that the maximum is taken over the set of projective measurements $\{\Pi_{k}\}$ \cite{4}.
\\By definition the conditional density operator $\rho^{AB}_{k}=\frac{1}{p_{k}}\{(I^{A}\otimes \Pi_{k}^{B})\rho^{AB}(I^{A}\otimes \Pi_{k}^{B})\}$ with $p_{k}=Tr[(I^{A}\otimes \Pi_{k}^{B})\rho^{AB}]$ as the probability of obtaining the outcome $k$, we can define the conditional entropy of $A$ as
$S(\rho^{AB}|\{\Pi_{k}\})=\sum_{k}p_{k}S(\rho_{k}^{A})$ with $\rho_{k}^{A}=Tr_{B}[\rho_{k}^{AB}]$ and $S(\rho_{k}^{A})=S(\rho_{k}^{AB})$.
It has been shown that $\textit{D}(\rho^{AB})\geq 0$ with the equal sign only for classical correlation \cite{5}.
\\Very recently, Rulli et al. \cite{3} have proposed a global measure of quantum discord based on a systematic extension of the bipartite quantum discord. Global quantum discord (GQD) which satisfy the basic requirements of a correlation function, for an arbitrary multipartite state $\rho^{A_{1}...A_{N}}$ under a set of local measurement $\{\Pi_{j}^{A_{1}}\otimes ... \otimes \Pi_{j}^{A_{N}}\}$ is defined as

\begin{eqnarray}
\textit{D}(\rho^{A_{1}...A_{N}})=\min_{\{\Pi_{k}\}}\,[S(\rho^{A_{1}...A_{N}}\|\Phi(\rho^{A_{1}...A_{N}}))\nonumber\\
-\sum_{j=1}^{N}S(\rho^{A_{j}}\|\Phi_{j}(\rho^{A_{j}}))].
\end{eqnarray}

Where $\Phi_{j}(\rho^{A_{j}})=\sum_{i}\Pi_{i}^{A_{j}}\rho^{A_{j}}\Pi_{i}^{A_{j}}$ and $\Phi(\rho^{A_{1}...A_{N}})=\sum_{k}\Pi_{k}\rho^{A_{1}...A_{N}}\Pi_{k}$ with $\Pi_{k}=\Pi_{j_{1}}^{A_{1}}\otimes ... \otimes\Pi_{j_{N}}^{A_{N}}$ and $k$ denoting the index string ($j_{1}...j_{N}$).
We could eliminate dependence on measurement by minimization the set of projectors $\{\Pi_{j_{1}}^{A_{1}}, ... ,\Pi_{j_{N}}^{A_{N}}\}$.
\\ By a set of von-Neumann measurements as

$$
\begin{array}{cc}
 \Pi_{1}^{A_{j}}=\left(\begin{array}{cc}
  \cos^{2}(\frac{\theta_{j}}{2})& e^{i\varphi_{j}}\cos(\frac{\theta_{j}}{2})\sin(\frac{\theta_{j}}{2}) \\
  e^{-i\varphi_{j}}\cos(\frac{\theta_{j}}{2})\sin(\frac{\theta_{j}}{2}) & \sin^{2}(\frac{\theta_{j}}{2}) \\
\end{array}
\right),
\end{array}
$$
$$
\begin{array}{cc}
 \Pi_{2}^{A_{j}}= \left(
\begin{array}{cc}
 \sin^{2}(\frac{\theta_{j}}{2})& -e^{-i\varphi_{j}}\cos(\frac{\theta_{j}}{2})\sin(\frac{\theta_{j}}{2}) \\
 -e^{i\varphi_{j}}\cos(\frac{\theta_{j}}{2})\sin(\frac{\theta_{j}}{2}) & \cos^{2}(\frac{\theta_{j}}{2})\\
\end{array}
\right),
\end{array}
$$

with $\theta_{j}\in[0,\pi)$ and $\varphi_{j}\in[0,2\pi)$ for $j=1,2,3$, the equation 10 reduces to

\begin{eqnarray}
\textit{D}(\rho(T))=\min_{\{\theta_{j},\varphi_{j}\}}\,[S(\rho(T)\|\Phi(\rho(T)))\nonumber\\
-\sum_{j=1}^{3}S(\rho^{A_{j}}\|\Phi_{j}(\rho^{A_{j}}))].
\end{eqnarray}

By tracing out two qubits, the one qubit density matrices representing the individual subsystems are

$$
\begin{array}{cc}
 \rho^{A_{j=1,2,3}}= \left(
 \begin{array}{cc}
 \rho_{11}+2\rho_{22}+\rho_{44}& 0 \\
 0 & \rho_{22}+2\rho_{44}+\rho_{88} \\
\end{array}
\right).
\end{array}
$$

To find the measurement bases that minimize quantum discord, after some algebraic calculation we have perceived that by adopting local
measurements in the $\sigma_{z}$ eigen basis for each particle, the value of quantum discord will be minimized. It leads to $S(\rho^{A_{j}}\|\Phi_{j}(\rho^{A_{j}}))=0$ and

\begin{eqnarray}
S(\Phi(\rho(T)))&=&-\{\rho_{11}\log_{2}(\rho_{11})+\rho_{88}\log_{2}(\rho_{88})\cr&+&
3\rho_{22}\log_{2}(\rho_{22})+3\rho_{44}\log_{2}(\rho_{44})\}.
\end{eqnarray}

The Entropy $S(\rho((T)))$ can be obtained as

\begin{eqnarray}
S(\rho(T))&=&-\{\rho_{11}\log_{2}(\rho_{11})+\rho_{88}\log_{2}(\rho_{88})\cr&+&2(\rho_{22}-\rho_{23})\log_{2}(\rho_{22}-\rho_{23})\cr
&+&(\rho_{22}+2\rho_{23})\log_{2}(\rho_{22}+2\rho_{23})\cr&+&
2(\rho_{44}-\rho_{46})\log_{2}(\rho_{44}-\rho_{46})\cr&+&(\rho_{44}+2\rho_{46})\log_{2}(\rho_{44}+2\rho_{46})\}.
\end{eqnarray}

In this case, the time evolution of quantum discord is explicitly obtained as

\begin{eqnarray}
\textit{D}(\rho(T))&=&-3\rho_{22}\log_{2}(\rho_{22})-3\rho_{44}\log_{2}(\rho_{44})\cr&+&2(\rho_{22}-\rho_{23})\log_{2}(\rho_{22}-\rho_{23})\cr
&+&(\rho_{22}+2\rho_{23})\log_{2}(\rho_{22}+2\rho_{23})\cr&+&
2(\rho_{44}-\rho_{46})\log_{2}(\rho_{44}-\rho_{46})\cr&+&(\rho_{44}+2\rho_{46})\log_{2}(\rho_{44}+2\rho_{46})\}.
\end{eqnarray}

%The responses of quantum discord to the diverse conditions can be seen from respective Figs.

\begin{figure}[htb]
\begin{center}
\includegraphics[width=1.2\linewidth]{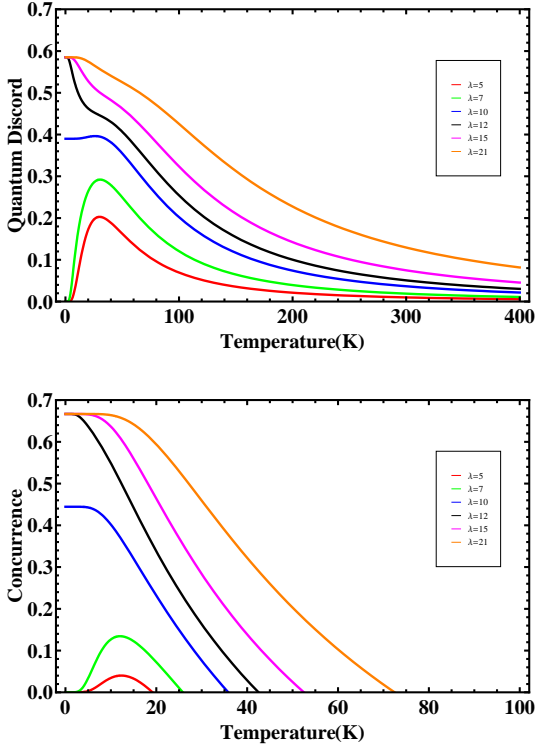}
\caption{(Color online) Discord(top panel) and concurrence(bottom panel) measures versus temperature for different values of F\"{o}rster interaction. Electric filed is zero.}
\label{fig1}
\end{center}
\end{figure}

\begin{figure}[htb]
\begin{center}
\includegraphics[width=1.2\linewidth]{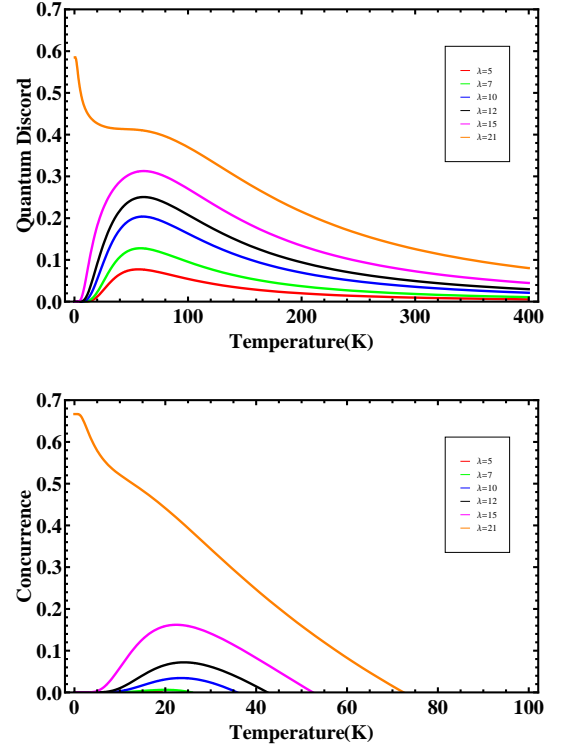}
\caption{(Color online) Discord(top panel) and concurrence(bottom panel) measures versus temperature for different values of F\"{o}rster interaction. The amount of applied Electric filed is about $20\times10^6$ $V/m$.}
\label{fig2}
\end{center}
\end{figure}

\section{3 Results and discussion}

In what follows, we theoretically test for the existence of quantum discord in aforementioned system containing excitonic qbits and that how it behaves with external electric field. From figures 1, comparing the concurrence and the discord measures at different temperatures shows that, quantum discord survives at relatively high temperatures where concurrence is zero, also figures depict at very low temperatures both measures yield the same result. At  higher temperatures, concurrence suddenly diminishes while quantum discord is still finite, slowly decreasing to zero. This is in accord with the fact that discord quantifies nonclassical correlations beyond entanglement. Generally, both quantum correlations
decay with temperature due to thermal relaxation effects. Figures demonstrate that when electrical bias is off the quantum correlations increase monotonically with increasing F\"{o}rster interaction. This can be explained in terms of the increasing  excitonic interaction that leads to increase the correlations. At F\"{o}rster interactions higher than about $10meV$ and at temperatures smaller than $20K$, concurrence is slightly larger than discord. In contrast, at lower F\"{o}rster interactions the opposite trend is observed and discord overcomes the concurrence almost for all temperatures that is because of as mentioned fact that discord indicates nonclassical correlations beyond entanglement even when $\lambda$ is low. From figures it is clear that at lower $\lambda$, Temperature helps to make the correlations so that for very low temperatures, till about $20K$(for discord) and $10K$(for concurrence), discord and concurrence increase with increasing temperature due to the thermal entanglement effects\cite{Arnesen}.\\
Figures 2 displays discord and correlation measures when $\hbar\Omega$ is assumed to be 2.5 $meV$ that corresponds to the electric field($E$) of about $20\times10^6$ $V/m$( this small value for electric field can be realized by experiments very easily). It is clear that discord and concurrence are smaller than that of observed in figure1. This indicates that  turning electric field on, decreases the correlations. It is because electric field makes all the dipoles align in the same direction that results in increasing dipolar repulsive interaction that immediately leads to the reduction of coulomb induced correlations\cite{Saeid}. For all the values of $\lambda$ the discord is survived with increasing $E$ while the concurrence dies for smaller $\lambda$. Our results shows that increasing the electric field overcomes the effect of F\"{o}rster interaction and removes the concurrence for all values of $\lambda$, however, nonzero discord can still be observed for higher values of electric field(figure3). This manifests the importance of discord to measure the quantum correlations.

\begin{figure}[htb]
\begin{center}
\includegraphics[width=1.2\linewidth]{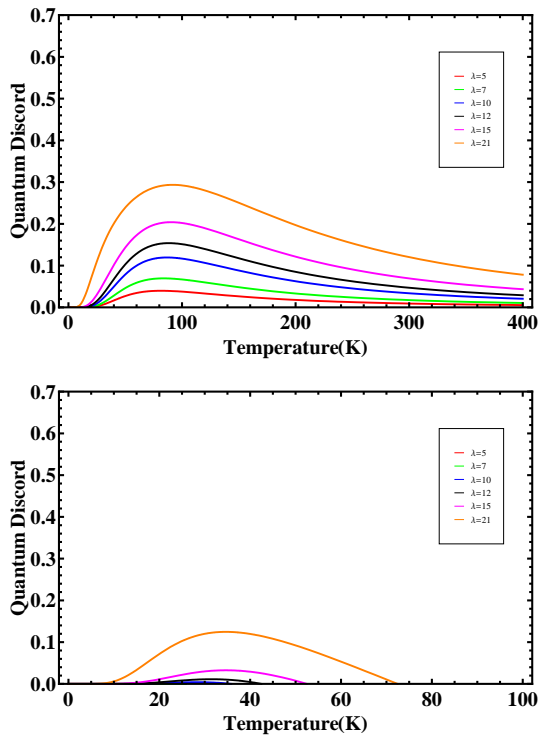}
\caption{(Color online) Discord(top panel) and concurrence(bottom panel) measures versus temperature for different values of F\"{o}rster interaction. The amount of applied Electric filed is about $40\times10^6$ $V/m$.}
\label{fig3}
\end{center}
\end{figure}

\section{4. Conclusions}

In summary, We studied the quantum discord and concurrence  measures in the array of optically driven coupled quantum dots. Qubits are excitons in each quantum dot that can be model by dipoles. We used the lower bound of concurrence for thermal density matrix of identical and equidistant coupled QDs at different temperatures. We found that the discord and concurrence are enhanced by increasing the parameter of  F\"{o}rster interaction that is resulted by increasing correlations with increasing this parameter. For very low temperatures and higher F\"{o}rster interaction,  concurrence is slightly larger than discord. In stark contrast, at low F\"{o}rster interactions the opposite trend is observed. Also in very low temperature region, because of thermal entanglement, both measures increase and then tend to zero for higher temperatures induced by thermal relaxation effects. We observed that switching electric field on, correlations diminish, however, discord is survived while concurrence dies for the all values of F\"{o}rster interaction. It is because electric field makes all the dipoles to be parallel that results in increasing dipolar repulsive interaction and finally decrease the entanglement.

\bibliographystyle{apsrev}

\end{document}